\documentclass[12pt,superscriptaddress,aps,prd,preprint]{revtex4}
\usepackage{amsmath}
\usepackage{amssymb}
\makeatletter
\newcommand{\bea}{\begin{eqnarray}}
\newcommand{\eea}{\end{eqnarray}}

\newcommand{\pa}{\partial}

\begin{document}
\title{The aether-modified gravity and the G\"{o}del metric}
\author{C. Furtado, J. R. Nascimento, A. Yu. Petrov}
\affiliation{Departamento de F\'{\i}sica, Universidade Federal da Para\'{\i}ba,\\
Caixa Postal 5008, 58051-970, Jo\~ao Pessoa, Para\'{\i}ba, Brazil}
\email{furtado,jroberto,petrov@fisica.ufpb.br}
\author{A. F. Santos}
\affiliation{Instituto de F\'{\i}sica, Universidade Federal de Mato Grosso,\\
78060-900, Cuiab\'{a}, Mato Grosso, Brazil}
\email{alesandroferreira@fisica.ufmt.br}

\begin{abstract}
We formulate the nonminimal aether-modified gravity whose action represents itself as a sum of the usual Einstein-Hilbert action and the CPT-even Lorentz-breaking aether-like gravity term proposed by Carroll. For this theory, we show that the G\"{o}del metric solves the modified Einstein equations under a corresponding modification of the matter, while, for the small Lorentz-breaking corrections, the matter turns out to be usual. On the other side, if the matter is suggested to be the same as in the usual general relativity, the G\"{o}del metric is modified by additive aether terms. 
\end{abstract}

\maketitle
Searching for the consistent gravity theory is, without doubts, the key problem of the actual theoretical physics. Beside of the more usual, Lorentz-invariant modifications which consist in adding the higher-derivative terms allowing for the improvement of the ultraviolet behaviour of the theories (see f.e. \cite{stelle}), the Lorentz and/or CPT-breaking modifications of gravity are intensively studied now. The most popular one among them is the Chern-Simons modified graviteither with an external Chern-Simons coefficient \cite{JaPi} or with a dynamical one \cite{Grumiller}, which breaks the CPT symmetry and, for a special form of the Chern-Simons coefficient, breaks also the Lorentz symmetry \cite{JaPi}. However, it is clear that the Chern-Simons modified theory does not exhaust the list of the possible Lorentz-breaking modifications of gravity. Many such modifications have been proposed in \cite{Kost}, and one of the interesting examples of such models is the Einstein-aether theory which has been very intensively studied in last years. Unlike the Chern-Simons modified gravity, it is CPT-even. A general review on it can be found in \cite{TJ, DM, Jacobson}. The most important results obtained for this theory are obtaining of its linearized wave spectrum \cite{Matt}, the development of its interpretation within the framework of metric affine  gravity \cite{Heinicke}, finding the constraints for its parameters from the observation of ultra-high energy cosmic rays \cite{Elliott}, studying of energy \cite{Eling}, discussion of its post-Newtonian parameters \cite{Foster}. The spherically symmetric black hole solutions in the Einstein-aether theory are examined in \cite{Ted}, its application to the consideration of the Lorentz breaking within the context of the inflationary expansion has been developed in \cite{Wise} (it is interesting to note that the similar cosmological behaviour has been found in \cite{Mota} for the model including spinor field), and to the study of the non-rotating neutron stars -- in \cite{Miller}. The instability of this theory is discussed in \cite{Donnelly}. Recently, some its applications have been performed in the Horava-Lifshitz gravity (for more information see f.e. \cite{Enrico, David}). These interesting results found for the Einstein-aether theory clearly motivate the search for other possible Lorentz-breaking CPT-even gravity extensions. A very interesting example of such extension has been proposed in \cite{Carroll} whose advantage consists in the fact of the simpler interaction of the constant Lorentz-breaking vector with the gravity which does not involve any covariant derivatives of this vector, and we will refer to this theory as to the nonminimal aether-modified gravity, in order to distinguish it from the Einstein-aether theory. Namely this theory is the main object of study in this paper. 

Within our study, we consider the consistency of the G\"{o}del metric \cite{Godel} within the framework of the nonminimal aether-modified gravity. This metric is the first known solution allowing for the closed timelike curves (CTCs). General aspects of the CTCs have been intensively discussed in \cite{Hawk}. Different issues related to the G\"{o}del metric have been studied in \cite{Godelworks,Reb}, where, in particular, it was shown to be consistent with the Chern-Simons modified gravity for the special form of the Chern-Simons coefficient. Also, we note that the compatibility of the G\"{o}del metric with the usual Einstein-aether theory have been discussed in \cite{Gurses}. Therefore, it is interesting to verify whether the Lorentz symmetry breaking in the form of the nonminimal aether modification is compatible with the G\"{o}del metric, in other words -- whether such a modification would admit the CTCs? This paper is devoted just to this problem.

To begin the study, let us briefly recall the main characteristics of the G\"{o}del metric \cite{Godel}
\bea
ds^2=a^2\Bigl[dt^2-dx^2+\frac{1}{2}e^{2x}dy^2-dz^2+2 e^x dt\,dy\Bigl],\label{godel}
\eea
where $a$ is a positive number. The corresponding nontrivial Christoffel symbols are 
\bea
\Gamma^0_{01}=1,\,\,\,\,\,\,\, \Gamma^0_{12}=\frac{1}{2}e^x,\,\,\,\,\,\,\,\Gamma^1_{02}=\frac{1}{2}e^x, \,\,\,\,\,\,\,\Gamma^1_{22}=\frac{1}{2}e^{2x},\,\,\,\,\,\,\,\Gamma^2_{01}=-e^{-x}.
\eea
The non-zero components of the Riemann tensor are
\bea
R_{0101}=-\frac{1}{2}a^2, \,\,\,R_{0112}=\frac{1}{2}a^2e^x, \,\,\, R_{0202}=-\frac{1}{4}a^2e^{2x}, \,\,\, R_{1212}=-\frac{3}{4}a^2e^{2x}.
\eea
The corresponding non-zero components of the Ricci tensor look like
\bea
\label{ricci}
 R_{00}=1, \,\,\,\,\,\,\,\, R_{02}=R_{20}=e^x, \,\,\,\,\,\,\,\, R_{22}=e^{2x}.
\eea
Finally, the Ricci scalar is
\bea
\label{scal}
R=\frac{1}{a^2}.
\eea
Following \cite{Carroll}, the action of the nonminimal aether-modified gravity looks like
\bea
S=A\int d^Dx\sqrt{|g|}(\frac{1}{2}R+\alpha_gu^au^bR_{ab}+L_{mat}),
\eea
with $A$ is a constant whose dimension and value depends on the dimension of the spacetime, $\alpha_g$ is a dimensionless parameter of aether coupling, $L_{mat}$ is a matter Lagrangian (to achieve the compatibility with the results of \cite{Godel}, we incorporate the cosmological constant into it), and $u^a$ is an background field introducing the Lorentz symmetry breaking (it is reasonable to impose the condition $u^au_a=1$, as in \cite{Carroll}, so, we break the Lorentz symmetry in a spontaneous way choosing this vector to be timelike,  $u^a=(\frac{1}{a},0,0,0)$, and, thus, $u_a=(a,0,ae^x,0)$). The corresponding equations of motion have the form \cite{Carroll}
\bea
\label{eqmot}
R_{ab}-\frac{1}{2}Rg_{ab}&=&\frac{\alpha_g}{2}\Big[R_{cd}u^cu^dg_{ab}+\\&+&\nabla_c\nabla_a(u^cu_b)+\nabla_c\nabla_b(u^cu_a)-\nabla_c\nabla_d(u^cu^d)g_{ab}-\nabla_c\nabla^c(u_au_b)\Big]+T_{ab},\nonumber
\eea
where $T_{ab}$ is an energy-momentum tensor of the matter involving the cosmological constant.
The nontrivial components of the Einstein tensor $G_{ab}=R_{ab}-\frac{1}{2}Rg_{ab}$ in our case look like
\bea
G_{00}=G_{11}=G_{33}=\frac{1}{2};\quad\, G_{02}=G_{20}=\frac{e^x}{2}, \quad\, G_{22}=\frac{3}{4}e^{2x}.
\eea 
The equations of motion are reduced to
\bea
\label{eqmot1}
G_{ab}=\frac{\alpha_g}{2}[R_{cd}u^cu^dg_{ab}+Z_{ab}]+T_{ab},
\eea
where the tensor $Z_{ab}$ is given by
\bea
Z_{ab}=\nabla_c\nabla_a(u^cu_b)+\nabla_c\nabla_b(u^cu_a)-\nabla_c\nabla_d(u^cu^d)g_{ab}-\nabla_c\nabla^c(u_au_b).
\eea
To find the components of $Z_{ab}$, we can use the facts that $\nabla_au^a=0$, and $\nabla_c\nabla_d(u^cu^d)=0$ for the G\"{o}del metric. Afterwards, we find that the only non-zero components of the $Z_{ab}$ are
\bea
Z_{00}=4, \quad\, Z_{11}=2, \quad\, Z_{22}=5e^{2x}, \quad\, Z_{02}=Z_{20}=4e^x.
\eea
Therefore we can write the system of Einstein equations:
\bea
\label{eqein}
(00) & & \frac{1}{2}=\frac{5\alpha_g}{2}+T_{00};\nonumber\\
(02) & & \frac{e^x}{2}=\frac{5\alpha_g}{2}e^x+T_{02};\nonumber\\
(11) & & \frac{1}{2}=\frac{\alpha_g}{2}+T_{11};\nonumber\\
(22) & & \frac{3}{4}e^{2x}=\frac{11\alpha_g}{4}e^{2x}+T_{22};\nonumber\\
(33) & & \frac{1}{2}=-\frac{\alpha_g}{2}+ T_{33}.
\eea
We see that the equations of motion are consistent for the energy-momentum tensor of the matter with the following nontrivial components:
\bea
&&T_{00}=\frac{1}{2}(1-5\alpha_g);\quad\, T_{02}=\frac{e^x}{2}(1-5\alpha_g); \quad\, T_{11}=\frac{1}{2}(1-\alpha_g);\nonumber\\
&& T_{22}=\frac{e^{2x}}{4}(3-11\alpha_g);\quad\, T_{33}=\frac{1}{2}(1+\alpha_g).
\eea
It is easy to see that this energy-momentum tensor differs from the usual one only by small terms proportional to $\alpha_g$.

{ Then, we can discuss a structure of the corresponding matter, using the methodology of \cite{Ishak} and suggesting that the matter is described as a relativistic fluid with the 4-vector of velocity $v^a$}: the density and pressure for the known $T_{ab}$, can be found as
\bea
\rho&=&T_{ab}v^av^b;\quad\, p=-\frac{1}{3}T_{ab}h^{ab}.
\eea
Here $h^{ab}=g^{ab}-v^av^b$ is a projecting operator (notice that our definitions differ from those ones used in \cite{Ishak} since we use an opposite signature). 

{ In the particular case when $v^a=u^a$, the density is} $\rho=\frac{1}{2a^2}(1-5\alpha_g)$, so, it is positive if the Lorentz breaking is enough small, $\alpha_g<\frac{1}{5}$. As for the pressure, it is equal to $p=\frac{3-\alpha_g}{6a^2}$, so, it is positive if $\alpha_g<3$ which is the weaker condition than the previous one. One notes that the results corresponding to the usual gravity are easily recovered. Also, it is easy to see that the interval $\alpha_g<\frac{1}{5}$ corresponds to the usual matter.

As a result, we can conclude turns out to be that if Lorentz breaking is enough small, the matter providing this compatibility is usual (neither ghost nor phantom one) although its energy-momentum tensor differs from that one in the case of the usual gravity by terms of the first order in the parameter of aether coupling $\alpha_g$. Another key conclusion is that the Lorentz symmetry breaking in the form of the nonminimal aether modification, for the usual (but modified following the prescriptions above) matter, at least for small Lorentz breaking does not make obstacles to the existence of the closed timelike curves (CTCs) whose existence is characteristic for the G\"{o}del metric. 

However, it is interesting whether other situations are possible. First, the natural question is -- since the aether is characterized by the four-vector of the velocity, whether the aether can replace the matter? In other words -- whether for the aether-modified gravity the G\"{o}del solution can arise for the empty space, for the appropriate choice of the aether velocity? It follows from the discussion above that the choice of the four-vector of the aether velocity in the form $u^a=(\frac{1}{a},0,0,0)$ does not allow for it. Indeed, the modified Einstein equations (\ref{eqein}) turn out to be inconsistent if we have either $T_{ab}=0$ or $T_{ab}=\Lambda g_{ab}$. In principle, it does not exclude the chance for existence of another form of aether velocity, we will study this scenario elsewhere.

At the same time, it is interesting to look for other solution, corresponding to the same matter as that one consider in the original paper \cite{Godel}. The essence of this method is as follows: we suggest that the metric solving the modified Einstein equations (\ref{eqmot1}) is an usual G\"{o}del metric $\bar{g}_{ab}$ (playing the role of the background metric within the approach of \cite{Velt}) and the variation $\alpha_g h_{ab}$ proportional to the small parameter $\alpha_g$ corresponding to the inclusion of the aether. In this case, the Ricci tensor and scalar curvature corresponding to the modified metric $g_{ab}=\bar{g}_{ab}+\alpha_g h_{ab}$ can be presented as power series in $\alpha_g$ up to the first order
\bea
&&\bar{R}_{ab}+\alpha_g \delta R_{ab}-\alpha_g\frac{1}{2}\bar{R}h_{ab}-\alpha_g\frac{1}{2}\delta R \bar{g}_{ab}-\frac{1}{2}\bar{R}\bar{g}_{ab}=\frac{\alpha_g}{2}[\bar{R}_{cd}u^cu^d\bar{g}_{ab}+\bar{Z}_{ab}]+\tilde{T}_{ab}+\nonumber\\&+&
\Lambda(\bar{g}_{ab}+\alpha_g h_{ab}),
\eea
where $\bar{R}_{ab}$ and $\bar{R}$ are the Ricci tensor and scalar curvature constructed on the base of the background metric $\bar{g}_{ab}$, and $\tilde{T}_{ab}=\rho u_au_b$ is a metric-independent part of the energy-momentum tensor. The first-order variations of the Ricci tensor and scalar curvature are presented here as $\alpha_g \delta R_{ab}$ and $\alpha_g\delta R$, with the $\alpha_g$ is shown explicitly.

Since the G\"{o}del metrics solves the unmodified Einstein equations $\bar{R}_{ab}-\frac{1}{2}\bar{R}\bar{g}_{ab}=\tilde{T}_{ab}+\Lambda \bar{g}_{ab}$, we conclude that the zeroth order in $\alpha_g$ of the equation above is automatically satisfied. Also, we took into account that for the G\"{o}del metric one has $\Lambda=-\frac{1}{2}\bar{R}$. So, the equation above is reduced to
\bea
\label{eqmot2}
\delta R_{ab}-\frac{1}{2}\delta R \bar{g}_{ab}=\frac{1}{2}[\bar{R}_{cd}u^cu^d\bar{g}_{ab}+\bar{Z}_{ab}].
\eea
From the \cite{Velt}, we can read off the results for $\delta R_{ab}$ and $\delta R$:
\bea
\delta R_{ab}&=&\frac{1}{2}[h^c_{c,ab}-h^c_{a,bc}-h^c_{b,ac}+h^{,c}_{ab,c}];\nonumber\\
\delta R&=&(h^{a,b}_{a,b}-h^{a,b}_{b,a}-\bar{R}^a_bh^b_a).
\eea
It allows to write down the equation (\ref{eqmot2}) in the closed form:
\bea
[h^c_{c,ab}-h^c_{a,bc}-h^c_{b,ac}+h^{,c}_{ab,c}]-
(h^{c,d}_{c,d}-h^{c,d}_{d,c}-\bar{R}^c_dh^d_c) \bar{g}_{ab}=\bar{R}_{cd}u^cu^d\bar{g}_{ab}+\bar{Z}_{ab}.
\eea
Substituting here the expressions (\ref{ricci},\ref{scal}), we can solve the system of equations for $h_{ab}$. To do it, we can suggest that all $h_{ab}$ depend only on $x_1=x$, just as the components of the G\"{o}del metric. For the simplicity, we can suggest that the tensor $h_{ab}$ has the same structure as the G\"{o}del metric, i.e. its non-zero component are $h_{00},h_{02}=h_{20},h_{11},h_{22},h_{33}$. In this case, we have the following system of equations:
\bea
\label{sys}
&&[h^c_{c,00}-2h^c_{0,0c}+h^{,c}_{00,c}]-
(h^{c,d}_{c,d}-h^{c,d}_{d,c}-\bar{R}^c_dh^d_c)a^2=5;\nonumber\\
&&[h^c_{c,02}-h^c_{0,2c}-h^c_{2,0c}+h^{,c}_{02,c}]-
(h^{c,d}_{c,d}-h^{c,d}_{d,c}-\bar{R}^c_dh^d_c) a^2e^x=5e^x;\nonumber\\
&&[h^c_{c,11}-2h^c_{1,1c}+h^{,c}_{11,c}]+
(h^{c,d}_{c,d}-h^{c,d}_{d,c}-\bar{R}^c_dh^d_c) a^2=1;\nonumber\\
&&[h^c_{c,22}-2h^c_{2,2c}+h^{,c}_{22,c}]-
(h^{c,d}_{c,d}-h^{c,d}_{d,c}-\bar{R}^c_dh^d_c) \frac{a^2}{2}e^{2x}=\frac{11}{2}e^{2x};\nonumber\\
&&
[h^c_{c,33}-2h^c_{3,3c}+h^{,c}_{33,c}]+(h^{c,d}_{c,d}-h^{c,d}_{d,c}-\bar{R}^c_dh^d_c)a^2=-1.
\eea
Since none of metric components depends on $x_3$, and there is no Christoffel symbols involving the index 3, the last equation of the system reduces to
\bea
h^{,c}_{33,c}+(h^{c,d}_{c,d}-h^{c,d}_{d,c}-\bar{R}^c_dh^d_c)a^2=-1
\eea

As a natural ansatz, we choose $h_{00}=A_1$, $h_{02}=h_{20}=A_2e^x$, $h_{11}=A_3$, $h_{22}=A_4e^{2x}$, with $A_1,A_2,A_3,A_4$ are constants (we note that this structure is completely compatible with the structure of the Einstein equations since each term in the corresponding equation will carry the same exponential factor, thus, the Einstein equations will be reduced to the algebraic ones). Since the $h_{33}$ is factorized from other variables, we can impose a condition $h^{,c}_{33,c}=B$ which is satisfied by $h_{33}=Ca^2e^{-x}-Ba^2x$. Indeed, if $h_{33}$ depends only on $x$ as other $h_{ab}$ do, we have $h^{,c}_{33,c}= g^{ab}(\pa_a\pa_bh_{33}-\Gamma_{ab}^c\pa_ch_{33})=g^{11}h_{33}^{\prime\prime}-g^{ab}\Gamma_{ab}^1h_{33}^{\prime}=
-\frac{1}{a^2}(h_{33}^{\prime\prime}+h_{33}^{\prime})=B$. It is worth to emphasize here that one cannot impose the condition $h_{33}=const$ since such a condition implies in the complete vanishing of the $h_{33}$ from the system (\ref{sys}), which, as a result, will be overflowed, including five equations for four variables, and thus inconsistent. As we will show further, the fact of nontrivial $x_1$-dependence of the $h_{33}$ implies in an essential modification of the metric which, however, does not preclude existence of the CTCs.
Therefore, we have
\bea
h^{c,d}_{c,d}-h^{c,d}_{d,c}-\bar{R}^c_dh^d_c=-\frac{1}{a^2}(1+B).
\eea

So, to mount the equations (\ref{sys}) we must obtain first (and then second) covariant derivatives of $h_{ab}$. The nontrivial ones look like
\bea
&&h_{01,0}=A_2-A_1,\quad\,  h_{12,0}=e^x(A_4-A_2-\frac{A_3}{2});\nonumber\\
&&h_{00,1}=-2(A_1-A_2),\quad\, h_{02,1}=(A_4-\frac{A_1}{2})e^x,\quad\, h_{22,1}=(2A_4-A_2)e^{2x};\nonumber\\
&&h_{01,2}=-\frac{e^x}{2}(A_1+A_3),\quad\, h_{12,2}=-\frac{e^{2x}}{2}(A_2+A_3).
\eea
It remains to find the second covariant derivatives of $h_{ab}$ and substitute them to the expression above. It is easy to see that among $h_{ab,cd}$ the only nontrivial terms are those ones which involve none or two indices 1, i.e. $h_{ij,kl}$, $h_{ij,11}$, $h_{11,ij}$, $h_{1i,j1}$, $h_{1i,1j}$, with indices $i,j$ take only values 0 and 2.

Then, $h_{c,d}^{c,d}=\nabla^d\nabla_d h$, where $h=h^c_c$ is a trace of the metric fluctuation. Then, $h=-\frac{1}{a^2}(A_1-4A_2+A_3+2A_4)-(Ce^{-x}-Bx)$, thus, $h_{c,d}^{c,d}=g^{33}h_{33,c}^{,c}=-\frac{B}{a^2}$. It remains to consider $h^c_{c,ab}=h_{,ab}$. To obtain a purely algebraic system for the coefficients $A_1,A_2,A_3,A_4,B$, it is natural to suggest that all terms in each of the equations of the system (\ref{sys}) are accompanied by the same exponential factor. However, while $h_{,00}=0$, we have $h_{,02}=-\frac{1}{2}(C+Be^x)$, $h_{,11}=-Ce^{-x}$, $h_{,22}=-\frac{e^{2x}}{2}(Ce^{-x}+B)$. Therefore we conclude that to obtain a purely algebraic system for $A_1,A_2,A_3,A_4,B$, one must have $C=0$ (we emphasize that this case is not an unique solution for $h_{ab}$, however, it represents itself as the simplest and hence most interesting solution).

The whole list of the nontrivial second derivatives of the metric fluctuation $h_{ab}$ looks like follows:
\bea
&&\nabla_0\nabla_0 h_{02}=\frac{e^x}{2}(A_1-A_2);\quad\
\nabla_0\nabla_2 h_{22}=-\frac{e^{3x}}{2}(2A_4-2A_2-A_3);\nonumber\\
&&\nabla_0\nabla_2 h_{00}=e^x(A_1-A_2);\quad\, \nabla_0\nabla_2 h_{02}=-\frac{e^{2x}}{4}(2A_4-2A_1-A_3);
\nonumber\\ &&
\nabla_0\nabla_2 h_{22}=-\frac{e^{3x}}{2}(2A_4-2A_2-A_3);\quad\,\nabla_0\nabla_0h_{22}=-e^{2x}(A_4-A_2-\frac{A_3}{2});\nonumber\\
&&\nabla_2\nabla_0 h_{00}=e^x(2A_1-2A_2);\quad\, \nabla_2\nabla_0 h_{02}=\frac{e^{2x}}{4}(-4A_4+3A_1+A_3);
\nonumber\\ &&
\nabla_2\nabla_0 h_{22}=-\frac{e^{3x}}{2}(4A_4-3A_2-A_3);\nonumber\\
&&\nabla_2\nabla_2 h_{00}=\frac{e^{2x}}{2}(3A_1-2A_2+A_3);\quad\, \nabla_2\nabla_2h_{02}=\frac{e^{3x}}{4}(2A_1+A_2+2A_3-2A_4);
\nonumber\
\eea
\bea
&&\nabla_2\nabla_2 h_{22}=-\frac{e^{4x}}{2}(2A_4-2A_2-A_3);\nonumber\\
&&\nabla_1\nabla_0 h_{10}=\frac{3}{2}A_1-3A_2-A_3+A_4;\quad\, \nabla_1\nabla_0 h_{12}=\frac{e^x}{2}(A_1-2A_2-A_3);
\nonumber\\ &&
\nabla_1\nabla_2 h_{10}=\frac{e^x}{2}(A_1-2A_2-A_3);\quad\,
\nabla_1\nabla_2 h_{12}=\frac{e^{2x}}{4}(A_1-2A_2-2A_3-2A_4);\nonumber\\
&&\nabla_0\nabla_1 h_{10}=2A_1-3A_2-\frac{1}{2}A_3+A_4;\quad\, \nabla_0\nabla_1 h_{12}=\frac{1}{2}(A_1-A_2)e^x;\nonumber\\
&&\nabla_2\nabla_1 h_{10}=\frac{e^x}{2}(3A_1-3A_2);\quad\,\nabla_2\nabla_1 h_{12}=\frac{e^{2x}}{4}(A_1+2A_2+A_3-4A_4);\nonumber\\
&&\nabla_1\nabla_1 h_{00}=3A_1-4A_2+2A_4;\quad\, \nabla_1\nabla_1 h_{02}=e^x(A_1-2A_2+2A_4);
\nonumber\\
&&\nabla_1\nabla_1 h_{22}=e^{2x}(\frac{1}{2}A_1-2A_2+3A_4);\quad\,
\nabla_0\nabla_0 h_{11}=2A_1-4A_2-A_3+2A_4;\nonumber\\
&& \nabla_0\nabla_2 h_{11}=e^x(A_1-A_2);
\quad\,
\nabla_2\nabla_0 h_{11}=(A_1-A_2)e^x;\quad\,\nabla_2\nabla_2 h_{11}=\frac{e^{2x}}{2}(A_1+A_3).\nonumber
\eea

Thus, the system of equations (\ref{sys}) is reduced to
\bea
\label{sys1}
&&[-2h^c_{0,0c}+h^{,c}_{00,c}]+B=4;\nonumber\\
&&[-\frac{B}{2}e^x-h^c_{0,2c}-h^c_{2,0c}+h^{,c}_{02,c}]+Be^x=4e^x;\nonumber\\
&&[-2h^c_{1,1c}+h^{,c}_{11,c}]-B=2;\nonumber\\
&&[-2h^c_{2,2c}+h^{,c}_{22,c}]=5e^{2x};\nonumber\\
&&h^{c,d}_{d,c}+\bar{R}^c_dh^d_c=\frac{1}{a^2}.
\eea
Also, one can see that $\bar{R}^c_dh^d_c=\frac{1}{a^4}A_1$.

Thus,
\bea
h^{c,d}_{d,c}=\frac{1}{a^4}(2A_2+A_3-2A_4).
\eea
It remains to solve the complete system of equations which looks like
\bea
&&(00):\, \frac{1}{a^2}(2A_1-A_2-2A_4)+B=4;\nonumber\\
&&(02):\, \frac{1}{a^2}(-A_3-2A_4)+\frac{B}{2}=4;\nonumber\\
&&(11):\,\frac{1}{a^2}(A_1-2A_4)-B=2;\nonumber\\
&&(22):\,\frac{1}{2a^2}(-A_1+A_2-A_3-5A_4)=5;\nonumber\\
&&(33):\, \frac{1}{a^2}(A_1+2A_2+A_3-2A_4)=1.
\eea
This system evidently has a nontrivial solution. Its explicit form looks like
\bea
A_1&=&-\frac{70}{51}a^2,\quad\,
A_2=-\frac{24}{17}a^2,\quad\,
A_3=\frac{43}{51}a^2,\quad\,
A_4=-\frac{37}{17}a^2,\quad\,
B=\frac{50}{51}.
\eea
As a result, we can write down the modified G\"{o}del metrics:
\bea
ds^2&=&a^2\Bigl[dt^2(1-\frac{70}{51}\alpha_g)-dx^2(1-\frac{43}{51}\alpha_g)+\frac{1}{2}e^{2x}dy^2(1-\frac{37}{34}\alpha_g)-
\nonumber\\&-&dz^2(1+\frac{50}{51}\alpha_g x)+ e^x dt\,dy(2-\frac{24}{17}\alpha_g)\Bigl].\label{godel2}
\eea
We see that this metric has an essential difference from the proper G\"{o}del metric (\ref{godel}) which manifests in an essential modification of the component $g_{33}$ acquired now the nontrivial $x$ dependence. In particular, it implies that this metric differs from the class of metrics considered in \cite{Reb}, so, under a corresponding rescaling of coordinates  and the scale factor $a$ (with the small value of $\alpha_g$) it reduces to 
\bea
ds^2&=&\tilde{a}^2\Bigl[dt^{\prime 2}-dx^{\prime 2}+\frac{1}{2}e^{2x^{\prime}}dy^{\prime 2}-
dz^{\prime 2}(1+Bx^{\prime})+ 2e^{x^{\prime}} dt^{\prime}\,dy^{\prime}\Bigl].\label{godel2}
\eea
Therefore, we must carry out the analysis of possibility of the CTCs within this metric.

It was shown in \cite{Godel} that CTCs exist if a number of properties of the space listed there is satisfied. So, let us verify validity of these properties for the new metric (\ref{godel2}). First of all, the metric component $g_{33}=-(1+B x)$ is not a constant more, and, it turns out to be also a nontrivial function of the $\phi$ angle. Thus, the CTCs cannot exist for this metric, so, the adding of the aether term excludes violation of the causality.  Second, the only exception is the plane $x_3=const$ where the causality can be violated \cite{Godel}. 

In this paper we have studied the compatibility of the G\"{o}del metric with the nonminimal aether-modified gravity. We have found that, first, to achieve the compatibility of the G\"{o}del metric in its own sense with this gravity model, one should modify the matter which, however, does not become an exotic one for the small Lorentz violation, second, the G\"{o}del metric is not consistent in the empty spacetime in the nonminimal aether-modified gravity as well as in the usual gravity, third, if one wants to preserve the same matter as in the usual case, that is, the same relativistic fluid as in \cite{Godel}, the G\"{o}del metric must be modified by additive of the aether term. We conclude that the causality is not violated in our Lorentz-breaking theory.

{\bf Acknowledgements.} Authors are grateful to  M. Reboucas for the interesting discussions. 
This work was partially supported by Conselho Nacional de
Desenvolvimento Cient\'\i fico e Tecnol\'ogico (CNPq). A. Yu. P. has been supported by the CNPq project No. 303438-2012/6.

\end{document}